# Synthesis and study of $\alpha$-Fe$_{1.4}$Ga$_{0.6}$O$_3$: An advanced Ferromagnetic Semiconductor

N. Naresh and R.N. Bhowmik[*]

Department of Physics, Pondicherry University, R. Venkataraman Nagar, Kalapet, Pondicherry-605014, India

[*]Corresponding author (RNB): Tel.: +91-9944064547; Fax: +91-413-2655734

E-mail: rnbhowmik.phy@pondiuni.edu.in

## Abstract

We report the synthesis of $\alpha$-Fe$_{1.4}$Ga$_{0.6}$O$_3$ compound and present its structural phase stability and interesting magnetic, dielectric and photo-absorption properties. In our work Ga doped $\alpha$-Fe$_2$O$_3$ samples are well stabilized in $\alpha$ phase (rhombohedral crystal structure with space group R3C). Properties of the present composition of Ga doped $\alpha$-Fe$_2$O$_3$ system are remarkably advanced in comparison with recently most studied FeGaO$_3$ composition. At room temperature the samples are typical soft ferromagnet, as well as direct band gap semiconductor. Dielectric study showed low dielectric loss in the samples with large enhancement of ac conductivity at higher frequencies. Optical absorption in the visible range has been enhanced by 4 to 5%. This composition has exhibited large scope of tailoring room temperature ferromagnetic moment and optical band gap by varying grain size and non-ambient (vacuum) heat treatment of the as prepared samples by mechanical alloying.

# 1. Introduction

Magnetic semiconductors with multiferroic properties belong to a special class of materials where spontaneous ferroelectric order (polarization) and ferromagnetic order n(magnetization) are coexisting. Recently, research interests on these materials are rapidly increasing considering their huge importance in basic sciences and room temperature applications in spintronic devices [1]. In order to search for spintronic materials, lots of efforts were devoted on conventional dilute ferromagnetic semiconductors. In dilute ferromagnetic semiconductors (DFMS) room temperature ferromagnetism was predicted [2] and also observed [3] by substituting a small amount of ferromagnetic atoms, e.g., Fe, Co, Ni, into lattices of conventional direct band gap semiconductors, e.g., ZnO, GaN, etc. However, observed ferromagnetic moment in DFMS in one hand small, and on other hand its origin is a matter of debate. In addition to that most of the DFMS materials are not good multiferroics [4], where one can control ferromagnetism by varying the electric field and vice versa. The mutual field dependent properties of the magnetic materials can be extremely useful in developing multifunctional materials, which can be applied on data storage medium, magnetic switching and biomedical applications [5].

Recently, there is an alternative proposal [6] to synthesize non-conventional ferromagnetic semiconductors for achieving room temperature ferromagnetism, semiconductor with moderate band gap and multiferroelectricity. In non-conventional ferromagnetic semiconductors, alloyed compound is formed by doping the suitable amount of direct band gap semiconductors, e.g., $TiO_2$, $Ga_2O_3$ and $Al_2O_3$, in the lattices of (antiferro/ferro) magnetic materials, e.g., $\alpha$-$Fe_2O_3$. Development of new magnetic semiconductors, based on the metal (Ti, Ga, Al) doped $\alpha$-$Fe_2O_3$, is a subject of world wide increasing research interest. The basic mechanism of the unusual magnetic, electric and electro-magnetic properties of metal doped $\alpha$-$Fe_2O_3$ (hematite) lies on the modifications of rhombohedral (111) planes of crystal structure. The planes of $Fe^{3+}$ ions are separated by layers of oxygen ($O^{2-}$) ions and adjacent planes form alternating ferromagnetic (FM) and antiferromagnetic (AFM) planes of $Fe^{3+}$ spins in $\alpha$-$Fe_2O_3$ crystal [7]. Formation of the metal doped $\alpha$-$Fe_2O_3$ alloy is possible due to the fact that metal ions $Ga^{3+}$ (0.62 Å), $Ti^{3+}/Ti^{4+}$ (0.64 Å) and $Al^{3+}$ (0.53 Å) have comparable ionic radii with $Fe^{3+}$ (0.67 Å) and stabilized in similar crystal structure. Attention on these materials is increasing due to their large scope of applications in the field of semiconductor based micro-electronics, opto-electronics and multifunctional devices [8, 9]. Interestingly, most of the dopants can induce states in the energy gap of $\alpha$-$Fe_2O_3$ and modified band structure can determine electro-magnetic properties of the compound [10-11]. Hence, metal doped binary magnetic oxides have all the potentiality to become next generation magnetoelectric (ME) materials for spintronics devices and sensor application.

Different research groups reported results on $\alpha$-$Fe_{2-x}Ti_xO_3$ series with main focus on the understanding of lamellar ferro/ferrimagnetism with large magnetic coercivity [12-14], spin glass and magnetic clustering [15], enhancement of electrical conductivity and band gap tailoring [10]; exchange bias [16] and half-metallic behaviour [14]. The charge mismatch at the interfaces of $Fe^{3+}$ and $Ti^{4+}$ and a disproportionate number of $Fe^{3+}$ and $Ti^{4+}$ ions can create multi-valent cations ($Fe^{3+}$, $Fe^{2+}$, $Ti^{4+}$ and $Ti^{3+}$) in $\alpha$-$Fe_{2-x}Ti_xO_3$ system [13, 14, 17] and leading to uncompensated ferrimagnetic moments, band gap tailoring, enhancement of conductivity and half-metallicity. Although $\alpha$-$Fe_{2-x}Ti_xO_3$ series exhibited reasonably large ferromagnetic moment [18, 19], but over all ferromagnetic moment at room temperature is not good enough for spintronic applications [20, 21]. Some of the composition, e.g., $FeTiO_3$ [22], may be promising candidate for electric-field switchable weak ferromagnetism, but multiferroelectricity (ferroelectric polarization) near to room temperature is small in these materials.

In comparison with $\alpha$-$Fe_{2-x}Ti_xO_3$ series, theoretical and experimental study of $\alpha$-$Fe_{2-x}Ga_xO_3$ series is limited. B. F. Levine et al. [23] showed a rich magnetic phase diagram for different Ga doping content. Recently, a solid solution of $\alpha$-$(Fe,Ga)_2O_3$ in the entire Fe/Ga was synthesized by J. M. G. Amores et al. [24], and their results showed the $\alpha$ phase stability only below 400 $^0$C. Experimental results of $\alpha$-$Fe_{2-x}Ga_xO_3$ series are encouraging to achieve the non-conventional ferromagnetic semiconductor for room temperature spintronic applications, including band gap tailoring, photoconductivity, ferromagnetism, multiferroic domain switching and ferroelectric polarization [25-28]. Although thin films of $\alpha$-$Fe_{2-x}Ga_xO_3$ series for different contents of Ga exhibited ferrimagnetism with wide range of $T_C$ (150-350 K) and possible magnetoelectric memories applications, it is found that most of the experimental works are confined mainly for the perovskite structured $FeGaO_3$ [29-33]. Unfortunately, Curie temperature ($T_C$) of $\alpha$-$FeGaO_3$ is well below of room temperature. Hence, suitable composition of $\alpha$-$Fe_{2-x}Ga_xO_3$ series must be selected for achieving room temperature ferromagnetic semiconductor.

This experimental work will focus on the study of $\alpha$-$Fe_{1.4}Ga_{0.6}O_3$, which seems to be a good non-conventional ferromagnetic semiconductor. Our main objective is to explore connectivity between structural phase stability and grain size dependent room temperature ferromagnetism, photoconductivity and band gap tailoring.

## 2. Experimental

Ga doped $\alpha$-$Fe_2O_3$ samples were prepared through mechanical alloying at atmospheric condition and subsequent heat treatment in non-ambient condition. Initially powders of Gallium Oxide ($\alpha$-$Ga_2O_3$, purity > 99.99 %) and Iron Oxide ($\alpha$-$Fe_2O_3$, purity > 99.99 %) were mixed in stoichiometric ratio to obtain the composition $\alpha$-$Fe_{1.4}Ga_{0.6}O_3$. The mixture was manually ground for 2 hours to prepare homogeneous mixture using agate mortar and pestle. This mixed material was taken into an 80 ml stainless steel bowl. Balls of Tungsten Carbide (5 mm) and steel (10 mm) were added to the mixture with the material to ball mass ratio 1:6. The bowl was fixed inside the FRITSCH (Pulverisette 6, Germany) planetary mono miller and mechanical alloying was carried out up to 100 hours milling time. The milling was stopped after every 2-3 hours interval for proper mixing of the alloyed powder as well as to minimize both local heat generation and agglomeration of alloyed particles. Small amount of milled powders was taken out after completing milling time 20, 60, 80 hours to check structural phase transformation during mechanical alloying and also to obtain samples with different grain size. For reference purpose, mechanical alloyed samples (without heat treatment) were denoted by MAX, where X = 20, 60, 80 and 100 for milling time 20 hrs, 60 hrs, 80, and 100 hrs, respectively. Room temperature X-ray diffraction (XRD) spectra of the samples were taken using Xpert Panalytical X-ray diffractometer with $CuK_{\alpha}$ radiation ($\lambda$=1.54056 $A^0$) in the 2$\theta$ range 20 to $80^0$ with step size $0.02^0$ and time per step 2 seconds. X-ray diffraction (XRD) spectra in Fig. 1a for MA20, MA100, MA100A8 samples showed that single phased crystal structure of α-$Fe_2O_3$ (Fig. 1b) is not formed in mechanical alloyed powders. Most of the XRD phase is consistent with α-$Fe_2O_3$ structure. A few additional peaks were noted for all milled samples and (*, #) marked in the spectra. We thought the single phased compound might be formed by high temperature heating of the milled samples. Under this assumption, we heated MA100 sample at $800^0C$ for 4 hours in air and thereafter, cooled the samples to room temperature @ 5 $^0C$/minute. Fig. 1a shows that this air-heated sample (MA100A8) is also single phased and additional peak (*) appeared in the room temperature spectrum. These additional peaks are not due to α-$Ga_2O_3$ (shown in Fig. 1a), but matching with β-$Ga_2O_3$ phase [24]. We understood that the additional phase may be due to oxidation of unreacted Ga atoms at grain boundaries. Then, pellet form of the MA100 sample was kept in non-ambient condition (vacuum ~ $10^{-6}$ mbar) inside the diffraction chamber of X-pert PANalytical X-ray diffractometer and XRD spectra were taken by heating the pellet at different temperatures. The vacuum heating time is typically 1 hour 40 minutes (i.e., XRD recording time only) at each temperature. The heating rate up to 800 $^0C$ and cooling to room temperature were maintained @ 30 $^0C$/ minute. The spectra of MA100 sample after non-ambient heating at different temperatures are shown in Fig. 1b. As temperature increases from 400 $^0C$ to 600 $^0C$ the height of extra peaks decreased significantly. We have found that single phased α-$Fe_2O_3$ structure is stabilized after non-ambient (vacuum) heating of MA100 samples at 800 $^0C$. Then XRD spectra of other as milled samples were recorded by directly heating them at 800 $^0C$ for 1 hour 40 minutes in vacuum condition. Although the samples were heated for 1 hour 40 minutes, but non-ambient heating at 800 $^0C$ for time less than 30 minutes is also sufficient for the structural stabilization of the compound in α-phase.

Hence, 800 $^0$C appears to be a critical temperature for the α-phase stabilization and also noted for chemical routed sample [24]. The as milled MA20, MA60, MA80 and MA100 samples after vacuum heat treatment at 800 $^0$C for 1 hour 40 minutes are denoted as MA20V8, MA60V8, MA80V8 and MA100V8, respectively (V8 means vacuum heat treatment at 800 $^0$C). In order to check the vacuum heat treatment effect on the grain size and physical properties of the samples, we continued mechanical milling of MA100V8 sample up to 200 hours (including preheated 100 hours milling time). These samples are denoted as V8M125, V8M160 and V8M200 for milling time 125, 160 and 200 hours, respectively. From elemental analysis of V8M200 sample using Bruker S4 pioneer wavelength dispersive X-ray fluorescence spectrometer (WDXRF), we have found the atomic ratio of Fe and Ga to 1.4:0.63. This is close to the expected value of 1.4:0.6 for the composition $Fe_{1.4}Ga_{0.6}O_3$.

Physical properties of the samples were studied at room temperature. Surface morphology of selected samples was studied using HITACHI S-3400N Scanning Electron Microscope. Elemental composition of the samples was detected using Energy dispersive X-ray (EDX) technique. Magnetization of the samples was measured using LakeShore 7404 vibrating sample magnetometer in the field range 0 to ± 15 kOe. Electrical ac conductivity of the samples was measured using Novocontrol broadband dielectric spectrometer. Optical absorption spectra of the samples were studied in the UV-Visible range.

**3. Results and Discussion**

Fig. 2 shows the XRD spectra of vacuum heat treated samples after cooling to room temperature. XRD profiles were fitted using FULLPROF program. For MA20V8 sample, one additional peak is visible at about $2\theta \sim 30^0$ due to the presence of minor $\beta$-$Ga_2O_3$ phase. Other wise, the profile of other vacuum heat treated samples were matched to the rhombohedral crystal structure with space group R3C of $\alpha$-$Fe_2O_3$ lattices. Grain size of the samples (as shown in Table 1) was estimated using Williamson-Hall method [34]. The grain size of the alloyed compound decreases with the increase of milling time. The vacuum heat treatment at 800 $^0$C after 100 hours milling has slowed down the grain size decreasing process with further increase of milling time up to 200 hours. At the same time, micro-strain in the crystal structure systematically increases with the increase of milling time, irrespective of in-between non-ambient heating of MA100 sample. It is interesting to note that lattice parameters (a and c) and cell volume (V) of the samples (shown in Table 1) have increased slowly up to the milling time 100 hours, despite the fact that radius of $Ga^{3+}$ (0.62 Å) is smaller than $Fe^{3+}$ (0.67 Å) [6]. Similar lattice expansion with decreasing grain size was observed in $\alpha$-$Fe_2O_3$, synthesized in identical mechanical milling [35]. The increase of cell parameters in the milled samples followed by vacuum heating at 800 $^0$C is primarily due to the effect of better alloying of two binary oxides ($\alpha$-$Ga_2O_3$ and $\alpha$-$Fe_2O_3$) to attain the structure of $\alpha$-$Fe_2O_3$ phase with higher cell parameters ($a = b$= 5.0386 Å, $c$ = 13.7498 Å, $V$ = 302.3 Å$^3$). Similar structural modification was also observed in the formation of mechanical milled $Cr_{1.4}Fe_{0.6}O_3$ alloy [34]. On the other hand, if the single phased MA100V8 compound is further mechanical milled the grain size as well as cell parameters both have decreased with the increase of milling time up to 200 hours. This means vacuum heat treatment has a significant impact on the stabilization of structural parameters also.

Fig. 3(a) shows magnetic field dependence of dc magnetization of the samples at room temperature. Strong room temperature ferromagnetism is exhibited in all samples. We can see that magnetic property of the $\alpha$-Fe$_{1.4}$Ga$_{0.6}$O$_3$ is well advanced in many ways and remarkably different from the low temperature ferromagnetism in recently most studied compound FeGaO$_3$ with perovskite structure (orthorhombic crystal with space group Pc2$_1$n) [29-33, 36]. Despite diamagnetic response of the used $\alpha$-Ga$_2$O$_3$ (data not shown in Fig. 3), room temperature ferromagnetic moment (spontaneous magnetization) in Ga doped $\alpha$-Fe$_2$O$_3$ structure has increased significantly (see Table 2). It is interesting to note that increase of the spontaneous magnetization (calculated from the linear extrapolation M(H > 10 kOe) data to the M axis at H = zero value) of the single phased alloy is strongly affected with the increase of milling time before vacuum heat treatment or in other words decrease of grain size. The ferromagnetic moment is more for 100 hours milled sample. This shows that increase of milling time provided better homogenization of Ga atoms into the proper lattice sites of Fe in $\alpha$-Fe$_2$O$_3$ structure before formation of single phased compound by vacuum heat treatment. The homogenization of Ga atoms into the lattice sites of Fe resulted in uncompensated ferromagnetic moment at the interfaces of rhombohedral planes, which is predicted as lamellar ferromagnetism [12-14]. Fig. 3(b) confirmed that doping of Ga atoms into the lattices of Fe during mechanical alloying process was enough to exhibit ferromagnetism in MA80 sample before non-ambient heat treatment, even though minor additional phase was coexisting in its crystal structure. Vacuum heat treatment of MA80 sample at 800 $^0$C not only provides a single phased structure, but room temperature ferromagnetism of the as alloyed sample, e.g., MA80, is also further enhanced in MA80V8. We have seen that milling of single phased MA100V8 slowly decreases ferromagnetic moment and the ferromagnetic moment loss is significant for 200 hours milled sample. We attribute this to increasing grain boundary disorder of the structurally stabilized magnetic grains [37]. From Fig. 3(c), we observe a symmetric ferromagnetic loop for $\alpha$-$Fe_2O_3$ sample. However, distorted (pinched shaped) ferromagnetic loop of Ga doped samples suggests the coexistence of two different magnetic layers, one is Fe rich and other one is Ga rich or mixture of both atoms in alternative layers of rhombohedral structure [12-14]. These distorted loops represent typical character of (layered) lamellar ferromagnetism and exchange bias effect or exchange spring magnet due to exchange coupling between two different magnetic layers in metal doped $\alpha$-$Fe_2O_3$ system [12-14, 30, 36]. The loops (inset of Fig. 3(b)) were used to calculate magnetic coercivity ($H_c$) of the samples. It may be noted (in Table 2) that coercivity of the vacuum heat treated samples is significantly low in comparison with the value (~ 1615 Oe) of $\alpha$-$Fe_2O_3$ sample. This shows that soft ferromagnetic character of the synthesized material increases by incorporating Ga atoms into $\alpha$-$Fe_2O_3$ structure. The pre-heated milling process has definitely affected the monotonic decrease of coercivity of the single phased samples and the coercivity attained minimum value for 100 hours milling. The data in Fig. 3(b) also showed that mechanical alloying process before vacuum heat treatment was useful to achieve better soft ferromagnet. Interestingly, coercivity and ferromagnetic moment both have increased in vacuum heat treated single phased samples. Ferromagnetic parameters of V8M160 and V8M200 samples suggest that by mechanical milling of single phased sample, e.g., MA100V8, coercivity can be tuned to higher value with out significant reduction of ferromagnetic moment. Similar increase of coercivity was also previously noted in mechanical milled $La_{0.67}Ca_{0.33}MnO_3$ nanoparticles [37].

Fig. 4 shows the frequency dependent ac conductivity data of selected samples, which were measured at different temperatures by applying ac field amplitude 1 V. Room temperature conductivity of the samples is very low, of the order $10^{-8}$-$10^{-11}$ S/cm. The increase of dc limit of the conductivity over the temperature scale 298 K- 473 K is nearly one order for MA20V8 and MA60V8 samples and nearly two orders for MA100V8 sample. However, frequency activated conductivity of the samples above 10 Hz has shown huge enhancement, of the order 5 to 6 at 10 MHz. Most remarkable feature is that higher frequency activated conductivity is weakly temperature dependent in all samples. This indicates wide band gap in the samples and charge carriers are strongly localized. At lower frequencies, samples exhibited a characteristic conductivity minimum and position of the minimum shifts to higher frequency with the increase of measurement temperature. This effect is not much pronounced for MA20V8 sample that showed incomplete single phased $\alpha$-$Fe_2O_3$ structure. In single phased samples (e.g., MA60V8 and MA100V8), the thermal activated shift of conductivity minimum is clearly observed (inset of Fig. 4 (c)) and one can expect such a competitive conductivity mechanism from Fe rich and Ga rich layers in these single phased $\alpha$-$Fe_2O_3$ structure. It is to be noted in Fig. 5 that dielectric loss of the samples is very low (2-0.002), even below 0.1 in the radio frequency range, and also in the range of reported values for $FeGaO_3$ [29]. Such materials can be considered as low loss components in electronic devices [38]. One notable feature is that MA20V8 sample exhibited relaxation (dielectric loss) peak (magnitude ~ 1) at 60 Hz and 10 Hz for the measurement temperature 25 $^0$C and 75 $^0$C, respectively. This means that dielectric loss peak shifts to lower frequency by increasing the measurement temperature and there is no dielectric loss peak at measurement temperature $\geq 100\ ^0C$ for MA20V8 sample. The dielectric loss peak magnitude in MA60V8 and MA100V8 samples is drastically reduced (nearly 10 times or more) in comparison with MA20V8 sample and traced at higher frequencies. The interesting feature is that dielectric loss peak of both MA60V8 and MA100V8 samples also shifts to lower frequencies with the increase of measurement temperature. Such shifting feature of dielectric loss peak with increasing measurement temperature is different from the conventional higher frequency shifting by thermal activated process, as observed in different materials [39-40]. Moreover, Table 3 shows that isothermal dielectric loss peak position appears at higher frequencies with the increase of milling time, while grain size decrease and better homogenization of Ga atoms in $\alpha$-Fe$_2$O$_3$ structure are taking place.

The doping of Ga atoms into the Fe sites of $\alpha$-Fe$_2$O$_3$ structure is further confirmed from the absorption spectra of the samples in UV-Visible wave length range (200 nm-1000 nm). One could see in Fig. 6a that the spectra of Ga doped $\alpha$-Fe$_2$O$_3$ samples are identical to the spectra of $\alpha$-Fe$_2$O$_3$ sample and drastically different from the spectrum of $\alpha$-Ga$_2$O$_3$ sample. The absorption band edge of $\alpha$-Ga$_2$O$_3$ is at about 300 nm and that of $\alpha$-Fe$_2$O$_3$ is at about 600 nm. Doping of Ga into $\alpha$-Fe$_2$O$_3$ is reflected in UV-Vis spectrum as the shift of absorption band towards higher wavelength side and the observations are consistent to chemical routed Ga doped $\alpha$-Fe$_2$O$_3$ sample [24]. The shift of the absorption band towards higher wavelength side indicates the decrease of energy band gap which is known as "Red-Shift" [41]. Hematite ($\alpha$-Fe$_2$O$_3$) shows a high intense light absorption peak at about 600 nm together with a less intense peak at about 630 and 860 nm, respectively. In the octahedral environment of $Fe^{3+}$ ions (i.e., $FeO_6$) five fold degeneracy of d orbital splits into low energy $t_{2g}$ ($d_{xy}$, $d_{xz}$, $d_{yz}$) and higher energy $e_g$ ($d_{x^2-y^2}$, $d_{z^2}$) levels [42]. Transition of electrons between these levels contributes to absorption bands in UV-Vis region. The absorption peaks observed in the present $\alpha$-$Fe_2O_3$ sample (Fig. 6 (a)) at ~860 nm, ~630 nm, ~540 nm, ~340 nm are corresponding to $^6A_1 \rightarrow {}^4A_2$, $^6A_1 \rightarrow {}^4A_1$, $^6A_1 \rightarrow {}^4E$, $6t_{14}\uparrow \rightarrow 2t_{2g}\downarrow$ transitions, respectively [43]. We have found that high intense absorption band (at about 600), as well as other absorption peaks are shifting towards higher wavelength side (Fig. 6 (b)) with the increase of milling time of Ga doped $\alpha$-$Fe_2O_3$ compound. Remarkably, there is an overall improvement of absorption of electromagnetic radiation over a wide spectrum of sunlight (visible range) and in near infra-red range after Ga doping in $\alpha$-$Fe_2O_3$ structure. This is the indication of improved photo-catalyst property of Ga doped $\alpha$-$Fe_2O_3$ samples. We noted typically 4.97% and 4.02% improvement of optical absorbption in V8M200 sample with respect to $\alpha$-$Fe_2O_3$ at the peaks 630 nm and 860 nm, respectively. From the absorption spectra it is evident that the samples are optically direct band gap semiconductors. In this case, absorption coefficient $\alpha(h\nu)$ is directly proportional to $(h\nu-E_g)^{1/2}/h\nu$, where $h\nu$ is the energy of the incident photon and $E_g$ is the optical band gap of the semiconductor [44]. The optical band gap ($E_g$) of the samples has been calculated using $(\alpha h\nu)^2$ vs. $h\nu$ plot (Fig. 6(c)). The intercept of the $(\alpha h\nu)^2$ vs. $h\nu$ straight line on the $h\nu$ axis provides $E_g$ of the sample [45, 46]. The calculated band gap energy of $\alpha$-$Ga_2O_3$ and $\alpha$-$Fe_2O_3$ are 4.61 eV and 2.08 eV, respectively are well compared with reported values [46]. The interesting feature is that band gap energy of the Ga doped samples is in between the band gap of $\alpha$-$Ga_2O_3$ (4.61 eV) and $\alpha$-$Fe_2O_3$ (2.08 eV), in other words more close to the band gap value of $\alpha$-$Fe_2O_3$.

We noted that band gap energy (~ 2.08 eV) of α-Fe$_2$O$_3$ has been widened by Ga doping (~ 2.48, 2.46, 2.41, 2.39 eV for MA20V8, MA60V8, MA100V8 and V8M200 samples, respectively) and the band gap modification is affected by grain size reduction and better homogenization of Ga atoms into the α-Fe$_2$O$_3$ structure. The results of Ga doped samples may be interesting for developing third generation solar cell materials [6, 10], where modulation of direct band gap semiconductor and increase of solar energy absorption in visible range are essential.

## 4. Conclusions

We have successfully synthesized Ga doped α-Fe$_2$O$_3$ in α phase, viz., α-Fe$_{1.4}$Ga$_{0.6}$O$_3$, using room temperature mechanical alloying of α-Ga$_2$O$_3$ and α-Fe$_2$O$_3$ powders and subsequent heating at 800 $^0$C in non-ambient (vacuum) condition. The samples are typical soft ferromagnet at room temperature, exhibiting hysteresis loop. Non-ambient heat treatment at 800 $^0$C of as milled samples provided not only the single phase crystal structure, but also enhanced room temperature ferromagnetism in the material. The results are interesting for practical applications of non-conventional ferromagnetic semiconductors and promised different aspects of tuning ferromagnetic parameters (spontaneous magnetization and coercivity) and semiconductor band gap by varying grain size of the material. Dielectric loss of the samples is extremely low due to direct band gap semiconductor nature of the material and can be used as low loss components in electronic devices.

## Acknowledgments

We thank to CIF, Pondicherry University, for providing material characterization facilities. RNB thanks to UGC for financial support [F.NO. 33-5/2007 (SR)].

**Figure Captions**

Fig.1 (a) XRD spectra of milled and air annealed at 840 $^0$C for 4 hours samples, (b) XRD spectra of MA100 sample recorded at different temperatures in non-ambient condition and compared to the XRD spectrum of $\alpha$-Fe$_2$O$_3$ sample.

Fig.2 (Colour online) XRD spectra of non-ambient heat treated $\alpha$-Fe$_{1.4}$Ga$_{0.6}$O$_3$ samples. * represents extra peak.

Fig. 3 (Colour online) (a) Initial M(H) curve of different samples.(b) M(H) loop of MA80 sample before and after non-ambient heat treatment. (c) M(H) loops and insert shows the magnified part of hysteresis loop.

Fig.4 (Colour online) ac conductivity vs. frequency of selected samples recorded at different temperature at 1 volt rms electric field. Insert shows thermal activated shift of ac conductivity minimum for MA100V8 sample.

Fig. 5 (Colour online) Dielectric loss of selected samples at different temperatures.

Fig.6 (Color online) (a, b) UV-Vis absorption spectra of different samples. (c) Plot of $(\alpha h\upsilon)^2$ vs photon energy (h$\upsilon$) to calculate E$_g$.

Table 1.

Grain size, lattice strain and cell parameters of different samples calculated using XRD data.

| Sample Name | Grain size (nm) | Lattice strain (%) | Cell parameter a (± 0.0005Å) | Cell parameter c (±0.001Å) | Cell volume V (±0.050(Å)$^3$) |
|---|---|---|---|---|---|
| MA20V8 | 47 | 0.06 | 5.0285 | 13.719 | 300.433 |
| MA60V8 | 39 | 0.10 | 5.0306 | 13.726 | 301.185 |
| MA80V8 | 30 | 0.16 | 5.0319 | 13.738 | 301.251 |
| MA100V8 | 24 | 0.18 | 5.0343 | 13.740 | 301.575 |
| V8M125 | 22 | 0.21 | 5.0279 | 13.716 | 300.280 |
| V8M160 | 21 | 0.25 | 5.0228 | 13.667 | 298.595 |
| V8M200 | 15 | 0.27 | 5.0189 | 13.681 | 298.447 |

Table 2.

Magnetic parameters coercivity ($H_C$), spontaneous magnetization ($M_S$) and optical band gap ($E_g$) of different samples calculated from M(H) curves.

| Sample Name | Coercivity (Oe) | $M_S$ (µB/Fe atom) |
|---|---|---|
| α-Fe$_2$O$_3$ | 1615 | 0.0034 |
| α-Ga$_2$O$_3$ | - | - |
| MA20V8 | 390 | 0.0135 |
| MA60V8 | 226 | 0.0181 |
| MA80V8 | 170 | 0.0228 |
| MA100V8 | 151 | 0.0243 |
| V8M125 | 212 | 0.0240 |
| V8M160 | 217 | 0.0236 |
| V8M200 | 211 | 0.0224 |

Table 3.

Frequency corresponds to dielectric loss peaks (peak in Tanδ vs frequency graph) at different measurement temperatures.

| Sample Name | Frequency corresponding to Tanδ peak (Hz) at | | | | | | | |
|---|---|---|---|---|---|---|---|---|
| | 25°C | 50°C | 75°C | 100°C | 125°C | 150°C | 175°C | 200°C |
| MA20V8 | 66 | 12 | - | - | - | - | - | - |
| MA60V8 | 34708 | 25557 | 11822 | 2646 | 2530 | - | - | - |
| MA100V8 | 44171 | 312371 | 24580 | 4599 | 726 | 382 | 352 | 273 |

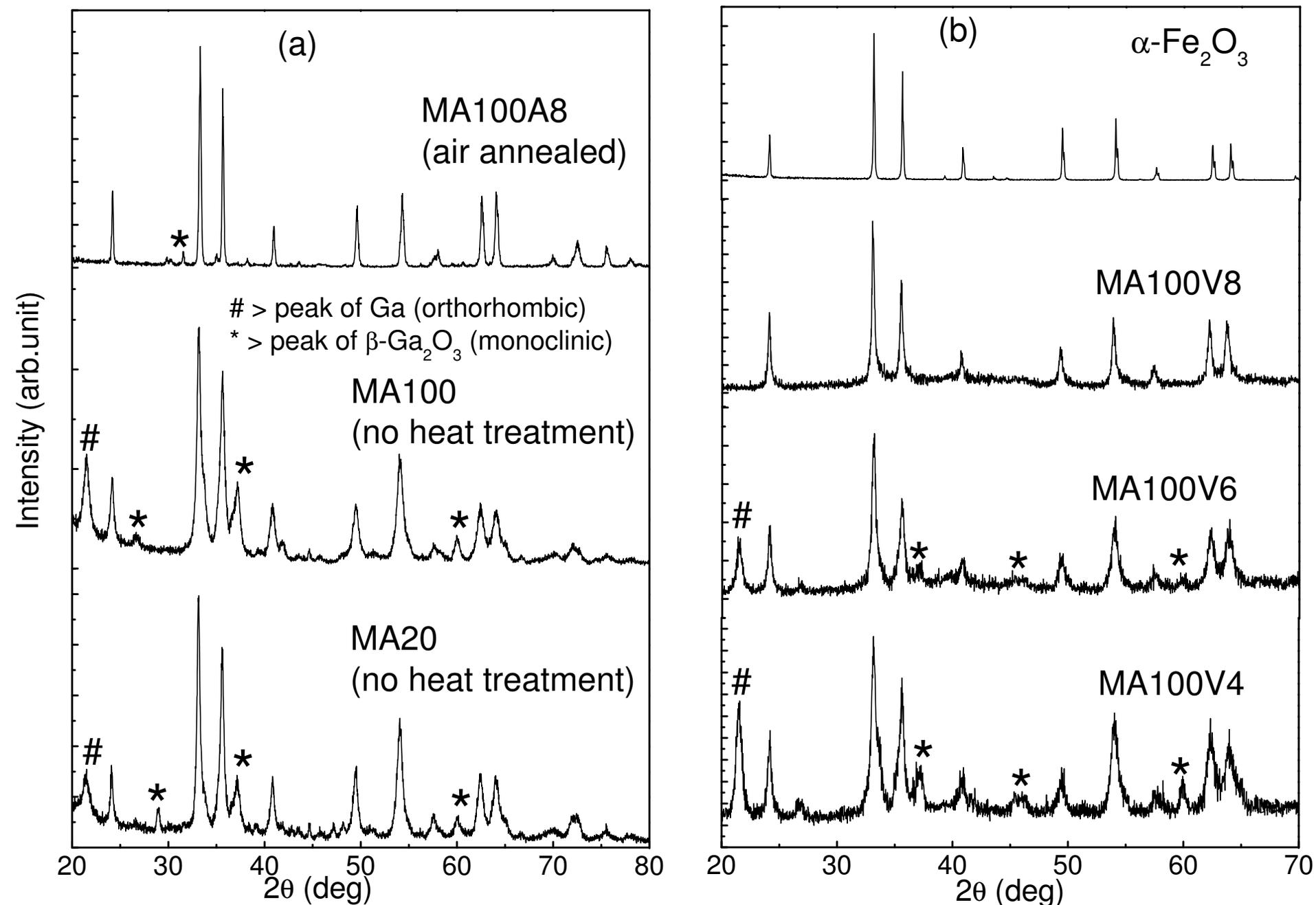

Fig.1 (a) XRD spectra of milled and air annealed at 840°C for 4 hours samples, (b) XRD spectra of MA100 sample recorded at different temperatures in non-ambient condition and compared to the XRD spectrum of $\alpha$-$Fe_2O_3$ sample.

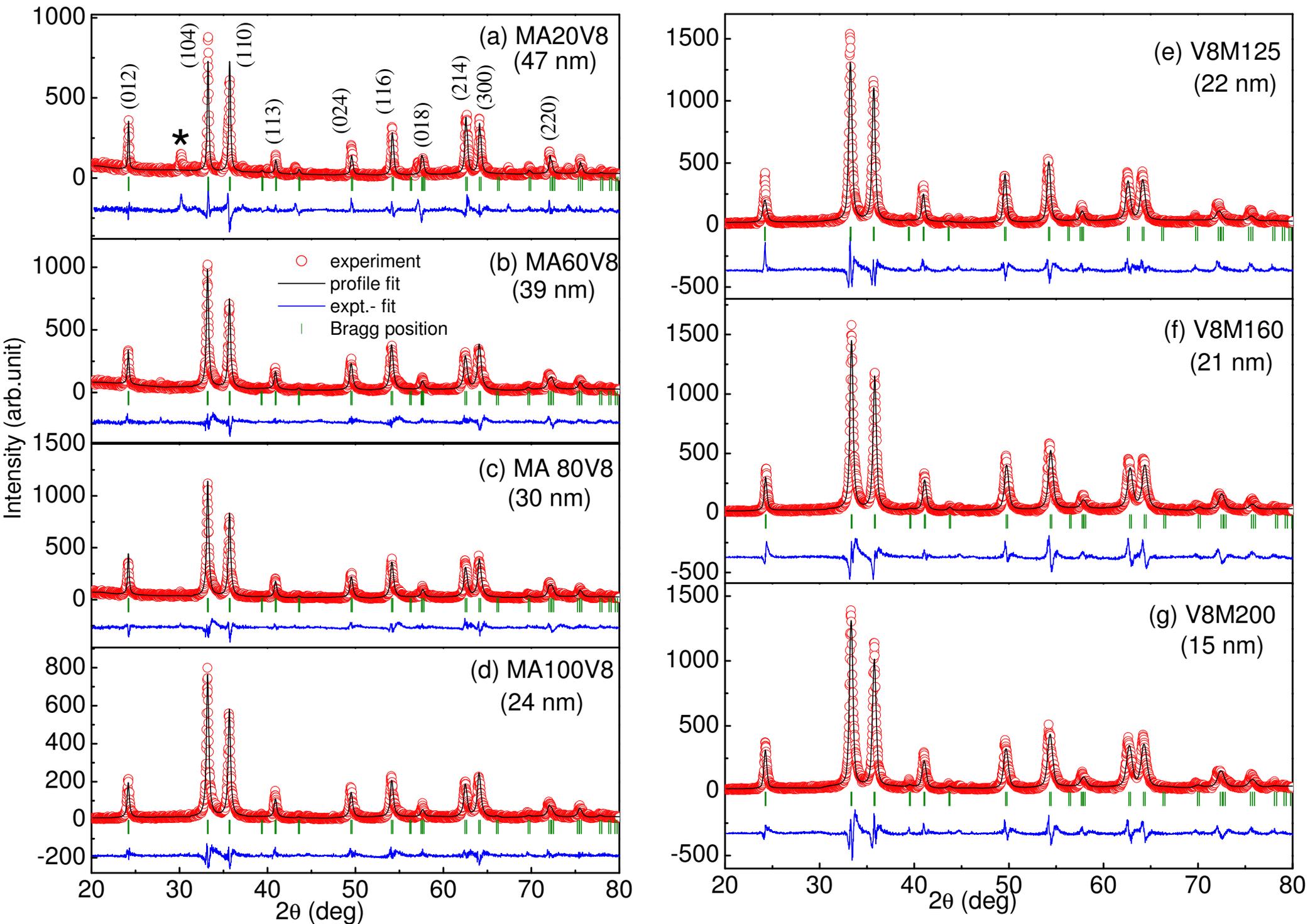

Fig.2 (Colour online) XRD spectra of non-ambient heat treated $\alpha$-$Fe_{1.4}Ga_{0.6}O_3$ samples. * represents extra peak.

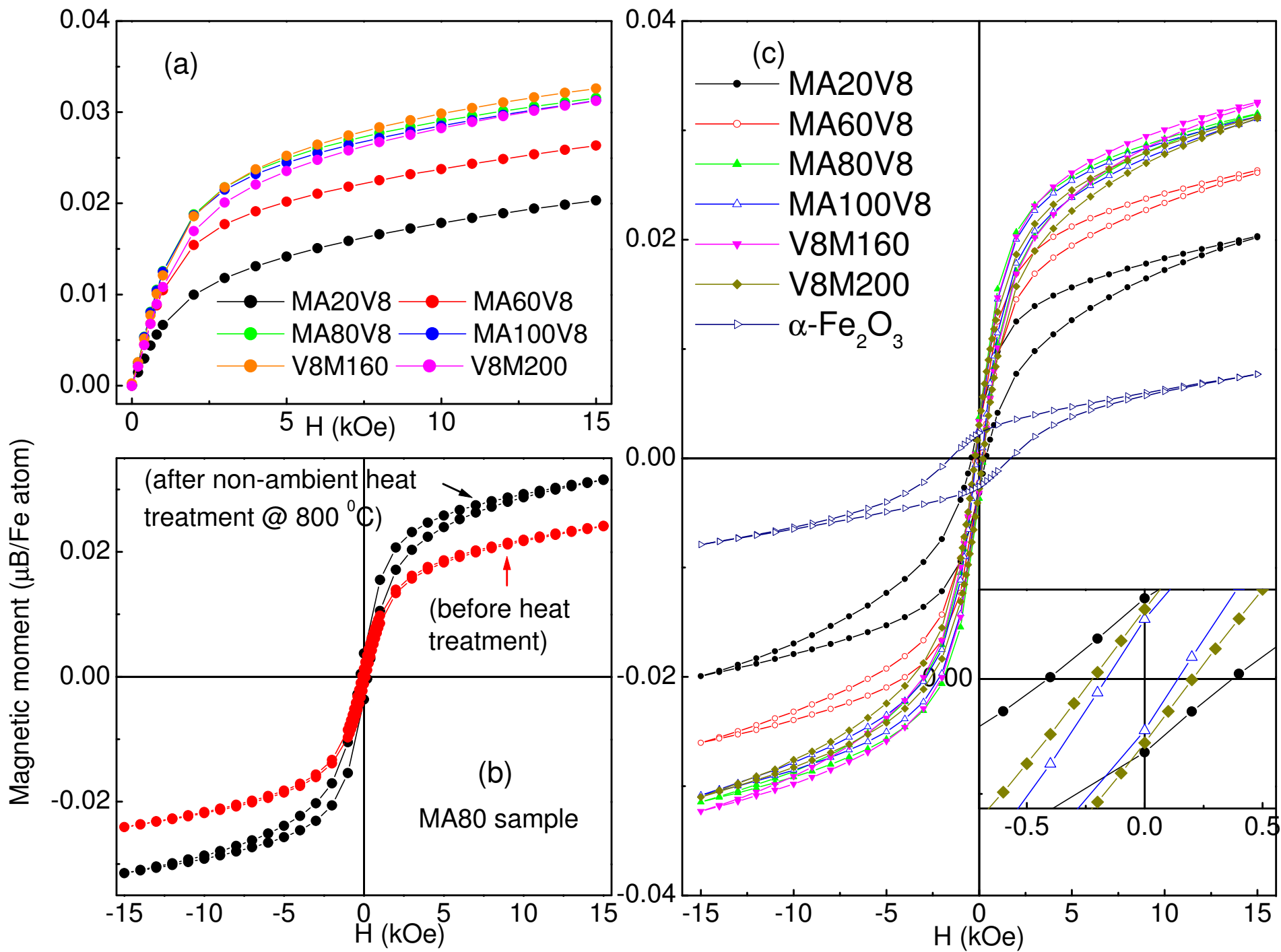

Fig. 3 (Colour online) (a) Initial M(H) curve of different samples. (b) M(H) loop of MA80 sample before and after non-ambient heat treatment. (c) M(H) loops and insert shows the magnified part of hysteresis loop.

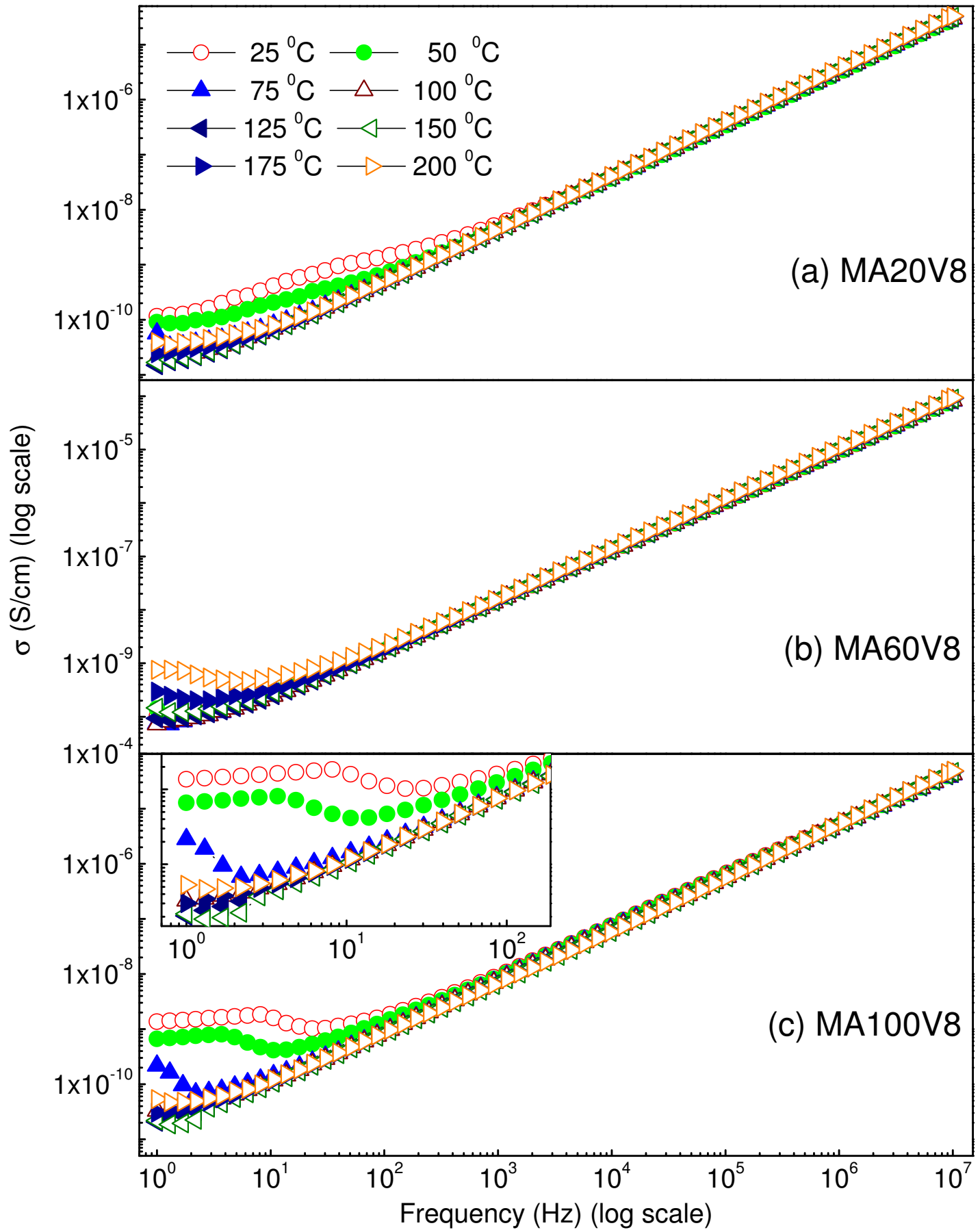

Fig.4 (Colour online) ac conductivity vs. frequency of selected samples recorded at different temperature at 1 volt rms electric field.
Insert shows thermal activated shift of ac conductivity minimum for MA100V8 sample.

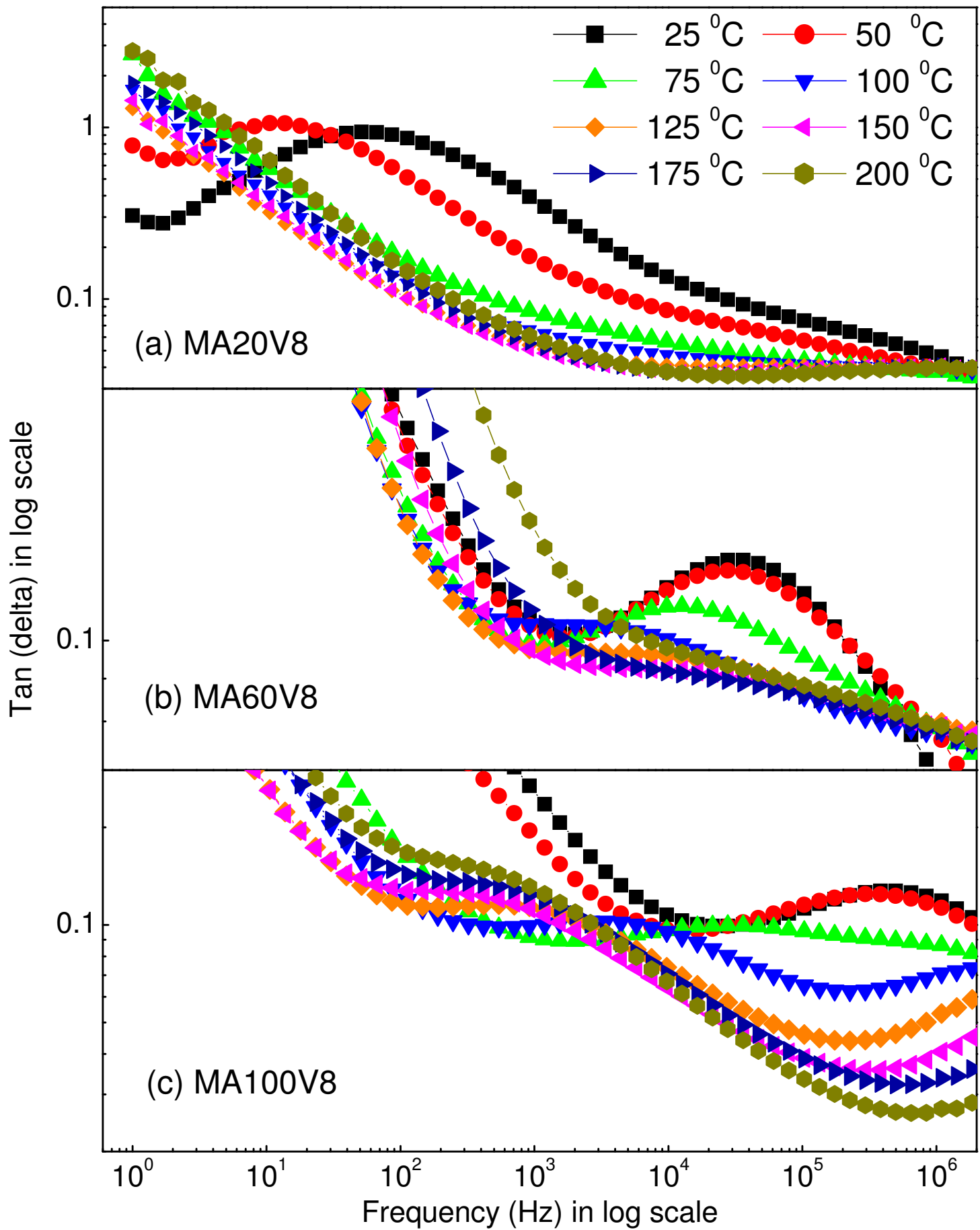

Fig. 5 (Colour online) Dielectric loss of selected samples at different temperatures.

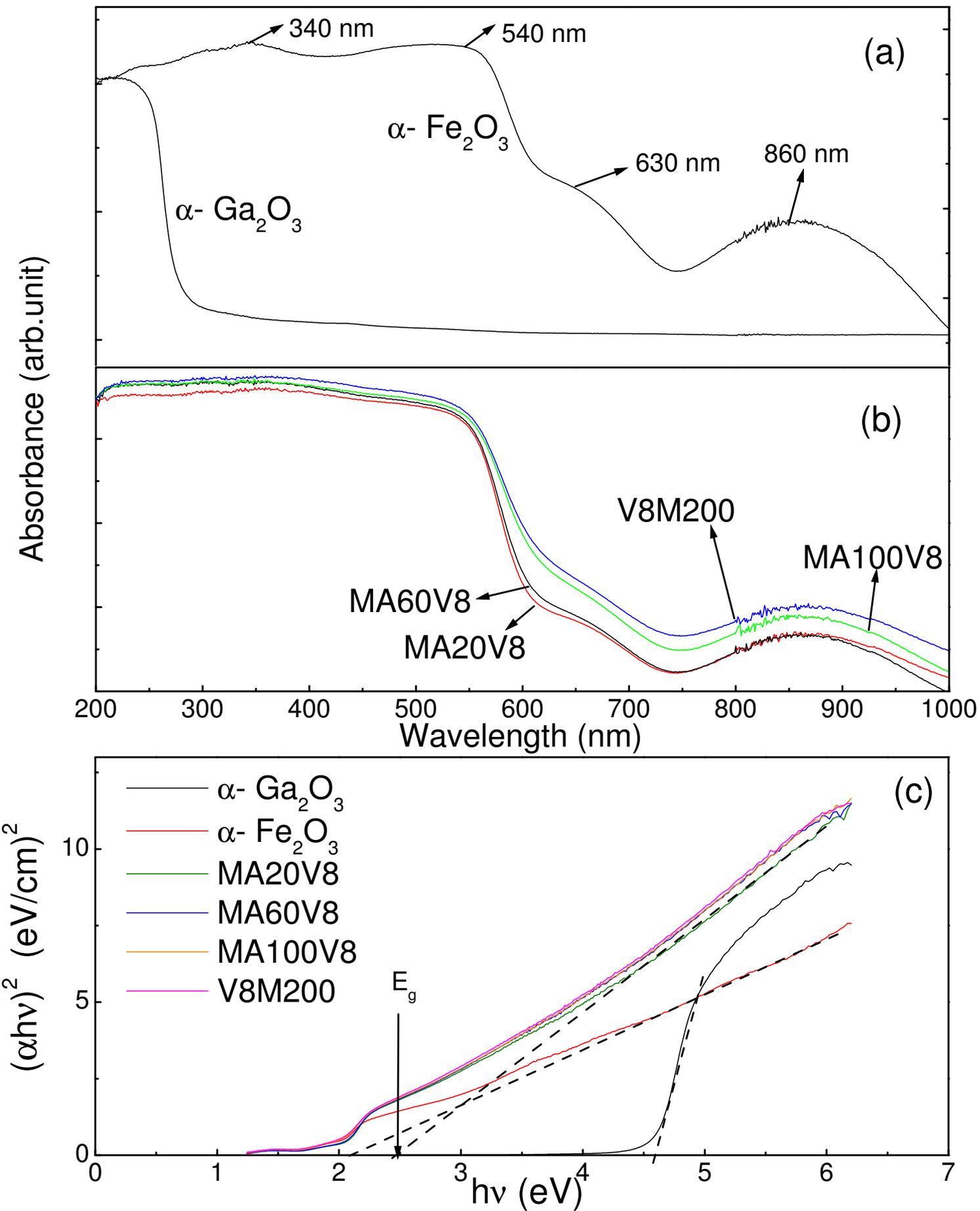

Fig.6 (Color online) (a,b) UV-Vis absorption spectra of different samples. (c) Plot of $(\alpha h\nu)^2$ vs photon energy ($h\nu$) to calculate $E_g$.